# Composition-dependent absorption of radiation in semiconducting MSi$_2$Z$_4$ Monolayers

Muhammad Sufyan Ramzan,[1*] Tomasz Woźniak,[2,3] Agnieszka Kuc,[3] Caterina Cocchi[1,4*]

[1]Institut für Physik, Carl von Ossietzky Universität, 26129 Oldenburg, Germany.

[2]Department of Semiconductor Materials Engineering, Wrocław University of Science and Technology, Wrocław, 50-370 Poland

[3]Helmholtz-Zentrum Dresden-Rossendorf, Center for Advanced System Understanding, CASUS, Untermarkt 20, 02826 Görlitz Germany.

[4]Center for Nanoscale Dynamics (CeNaD), Carl von Ossietzky Universität, 26129, Oldenburg, Germany.

Correspondence to: muhammad.sufyan.ramzan@uni-oldenburg.de,

caterina.cocchi@uni-oldenburg.de

# Abstract

The recent synthesis of MoSi$_2$N$_4$ material, along with theoretical predictions encompassing the entire family of chemical analogs, has opened up a new array of low-dimensional materials for a diverse range of optoelectronics and photovoltaics applications. In this study, we conducted state-of-the-art many-body first-principles calculations to analyze the quasi-particle electronic structure of the material class MSi$_2$Z$_4$ (where M = Mo, W, and Z = N, P, As, Sb). All monolayers display a direct band gap at the K point, with the exception of MoSi$_2$N$_4$. In tungsten-based compounds, the fundamental-gap can be adjusted over a significantly broader energy range compared to their molybdenum-based counterparts. Additionally, increasing atomic weight of the Z, both the band gap and exciton binding energies decrease. A noteworthy feature is the absence of a lateral valley (Λ or Q) near the conduction band minimum, indicating potential higher photoluminescence efficiencies compared to conventional transition-metal dichalcogenide monolayers. The optical spectra of these materials are predominantly characterized by tightly bound excitons, leading to an absorption onset in the visible range (for N-based) and in the infrared region (for others). This diversity offers promising opportunities to incorporate these materials and their heterostructures into optoelectronic devices, with tandem solar cells being particularly promising.

## 1.1  Introduction

Graphene was successfully exfoliated from its bulk form in 2004,[1] sparking a surge of interest in other two-dimensional (2D) materials, which offer a variation of electronic properties.[2–11] The current catalogue of 2D materials offers a diverse range of insulating, semiconducting, semi-metallic, and metallic monolayers (MLs).[12–17] Among the reported 2D materials, transition metal dichalcogenides (TMDCs) have been extensively studied due to their direct band gaps, high carrier mobilities, and stability in ambient conditions.[4,18,19] However, their optoelectronic properties are limited by the presence of a lateral valley, so-called Λ or Q, near the conduction band minimum (CBM), which provides non-radiative recombination sites for the excited carriers, thus, reducing the photoluminescence efficiency.[20–23] Several methods have been proposed to suppress this non-radiative recombination channel and, thus, enhance the quantum yield of TMDC MLs,[22,24,25] but despite these efforts, additional research is needed to make them ready for commercial optoelectronic applications. One possible strategy to overcome current limitations of these materials is to interface them with other organic and inorganic semiconductors with smaller band gap to form so-called "tandem stacks" that lead to improved solar cell efficiency by means of photon upconversion.[26,27] However, despite the efforts devoted to improve the performance of TMDCs, the search for other 2D semiconductors is still an active area of research.

Recently, a new 2D material, $MoSi_2N_4$, with a crystal structure analogous to TMDCs, was synthesized using chemical vapor deposition method[28]. It belongs to $P\bar{6}m2$ space group and has a thickness of seven atomic planes with $MoN_2$ layer sandwiched between two layers of SiN, see **Figure 1**. $MoSi_2N_4$ has an indirect bandgap of 1.94 eV and exhibits excitonic transitions at 2.21 eV (the so-called A resonance) and 2.35 eV (B resonance) originating from the spin-splitting of the valence band maximum (VBM), similar to TMDCs.[29,30,28] Moreover, $MoSi_2N_4$ has high electron (270 cm$^2$ V$^{-1}$s$^{-1}$) and hole (1200 cm$^2$ V$^{-1}$s$^{-1}$) mobilities resulting in a high on-off ratio of 4000 at 77 K in a field-effect transistor.[28] A recent theoretical study has demonstrated that its band edges are well protected by the local environment.[31] Moreover, density functional theory calculations predicted an entire class of 2D analogs, with a general formula of $MA_2Z_4$, where M indicates Group-2 or transition metals, A stands for elements belonging to Group-13 or -14, and Z species of Group-15 or -16. Similar to TMDCs, different structural phases of MSi$_2$Z$_4$ have also been proposed and investigated.[32]

So far, numerous ground-state calculations for several members of this new material class have been reported.[28,33–36] Most of these studies showed that MA$_2$Z$_4$ MLs have direct band gaps at the high-symmetry point K, with the exception of MoSi$_2$N$_4$ and WSi$_2$N$_4$ that both exhibit indirect band gaps with VBM at Γ point. It is worth highlighting that, unlike the TMDCs, MSi$_2$Z$_4$ do not have the Λ valley near the CBM, which is responsible for detrimental electron-hole recombination in TMDCs, as discussed above. The absence of the Λ valley between Γ and K suggests these materials as potential candidates for optoelectronics and photovoltaics. However, a detailed investigation of the electronic and optical characteristics of these systems based on state-of-the-art *ab initio* methods is necessary to substantiate this intuitive claim. To date only a few reliable studies of this kind are available in the literature.[31,37–39]

In this work, we present a systematic study of the electronic and excitonic properties of MSi$_2$Z$_4$ (M = Mo, W and Z = N, P, As, and Sb) family of MLs carried out in the framework of density functional theory (DFT) and many-body perturbation theory (GW approximation and Bethe-Salpeter equation). We analyze the trends obtained at varying composition for the electronic and optical band gaps as well as for the excitons and their binding energies, identifying trends that enable engineering of these materials and their properties to maximize their optical performance. We find that the optical onset of the N-based MLs is in the visible region, while for the others, it is shifted to the infra-red (IR), suggesting intriguing perspectives for efficient tandem solar cells based on vertically stacked heterostructures of these systems.

## 2.1 Results and Discussion

### i. Structural properties

MSi$_2$Z$_4$ ML systems have a hexagonal crystal structure belonging to the D$_{3h}$ point group. Its unit cell includes seven atomic layers, with MZ$_2$ layer sandwiched between two Si-Z layers (see **Figure 1**a, b). The optimized in-plane lattice constants *a* = *b* (see **Table 1**) are sensitive to the Z elements and increase monotonically as their atomic mass increases. Similar to the TMDCs in the 2H phase, MLs with the same Z element, but different metal atom M, have nearly identical lattice constants, such that, e.g., $MoSi_2P_4$ and $WSi_2P_4$ have the same lattice constant equal to 3.454 Å. This interesting feature promises the creation of Mo- and W- based heterostructures without lattice mismatch. Our calculated lattice constants agree well with earlier experimental and theoretical reports[31,36,38]. All the MLs are predicted to be mechanically stable[28,31,36]. The first Brillouin zone is hexagonal (see **Figure 1**c) with the usual high symmetry points ±K indicating valleys with opposite polarization (**Figure 1**d).

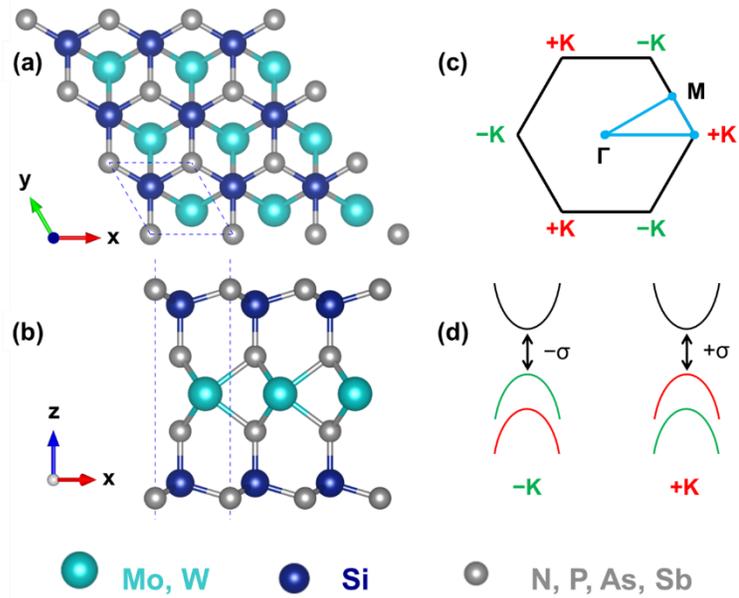

**Figure 1.** a - b Top and side views of the $MSi_2Z_4$ ML atomic structure with the unit cell boundaries maked by dashed lines. (c) Bruillion zone with the path connecting high symmtery points indicated in light blue. (d) Schematic depiction of ±K valley polarization.

## ii.  Electronic properties

The band structures of the considered $MSi_2Z_4$ monolayers are shown in **Figure 2**. These results, obtained at the PBE level including SOC, exhibit an underestimated band gap, due to the known shortcomings of this semi-local functional. Yet, the qualitative picture provided by these results is consistent with the one delivered by more sophisticated methods such as GW, see Figure S7 and S8. The VBM and CBM of all monolayers are situated at the K point, thus giving rise to direct band gaps, except for MoSi$_2$N$_4$ and WSi$_2$N$_4$ which have an indirect band gap with the VBM at the Γ point. In these materials, the highest valence band has a maximum at the K point 234 meV (for MoSi$_2$N$_4$) and 50 meV (for WSi$_2$N$_4$) below the VBM. In all other monolayers, the K valley is always higher than the Γ valley and their difference increases for heavier Z elements. We also notice a dependence of the SOC splitting of the valence and conduction bands on the composition of the MLs. In particular, heavier Z elements increase the VBM splitting, leading to an overall decrease of the band gap. The conduction-band splitting is only noticeable for MSi$_2$As$_4$ and MSi$_2$Sb$_4$, but it is still below 35 meV even for the latter system including the heavier element Sb. The corresponding values of band gaps and spin splitting are given in **Table 1**. The geometrical and electronic parameters and their trends agree with previous studies[31,32,36].

**Table 1.** Optimized lattice constants of the hexagonal unit cells of the MSi$_2$Z$_4$ MLs, *a* = *b*, layer thickness *d*, SOC-induced splitting of the highest valence band (v) and lowest conduction band (c) at the K point, PBE and GW gap (direct gap in parenthesis if the fundamental gap is indirect), as well as optical band gap corresponding to the lowest bright excitation predicted by the BSE, and its binding energy (B.E.).

| System | d (Å) | *a* = *b* (Å) | SOC-splitting of v/c (meV) | Band gap (eV) PBE | Band gap (eV) GW | Band gap (eV) Optical | B.E. (eV) |
|---|---|---|---|---|---|---|---|
| MoSi$_2$N$_4$ | 7.0 | 2.900 | 130/3 | 1.781 (2.015) | 2.720 (2.859) | 2.470 | 0.389 |
| MoSi$_2$P$_4$ | 9.4 | 3.454 | 138/4 | 0.620 | 1.067 | 0.859 | 0.208 |
| MoSi$_2$As$_4$ | 9.9 | 3.597 | 181/16 | 0.528 | 0.881 | 0.693 | 0.188 |
| MoSi$_2$Sb$_4$ | 10.9 | 3.879 | 226/25 | 0.263 | 0.495 | 0.380 | 0.115 |
| WSi$_2$N$_4$ | 7.0 | 2.889 | 400/10 | 2.110 (2.160) | 3.047 | 2.624 | 0.423 |
| WSi$_2$P$_4$ | 9.4 | 3.454 | 439/6 | 0.300 | 0.652 | 0.452 | 0.200 |
| WSi$_2$As$_4$ | 9.9 | 3.599 | 503/25 | 0.211 | 0.467 | 0.291 | 0.176 |
| WSi$_2$Sb$_4$ | 10.9 | 3.884 | 510/19 | 0.031 | 0.178 | 0.019 | 0.159 |

The most striking feature in the electronic structures of $MSi_2Z_4$ is the absence of the Λ valley between the high-symmetry points Γ and K near the CBM, in contrast to conventional TMDCs,[18,19,40,41] see Figure 2 and S1. It is worth noting that a feature analogous to the Λ valley is still present in the band structure of the $MSi_2N_4$ (see Figure 2), but it is energetically much higher than the CBM at K compared to the Λ-valley in the TMDCs. The residual presence of this band minimum in the conduction region of the $MSi_2N_4$ can be explained by the largely different electronegativities of Si (1.8) and N (3.0) compared to Si and the heavier Z elements. For the same reason, an additional valley also appears at the M point, which is dominated by the Si and N p orbitals (see Fig. S5 in ref. [36]). For W-based MLs, these two valleys are closer to the CBM than in the Mo-based ones. In all other MLs, with heavier Z elements than N, the Λ and M valleys disappear and a new valley appears (CBM+1) above the CBM, composed of *p* and *d* orbitals of the Z and M atoms, respectively.[36] Furthermore, with heavier Z elements, the spin-splitting of CBM+1 valley increases and the energy difference between CBM and CBM+1 valley at K decreases. The Λ valley, when energetically close to the K valley, provides nonradiative recombination channels, and hence, suppresses the photoluminescence quantum yield.[42,43,41,25] Due to the absence of this feature in monolayers $MSi_2Z_4$, excellent optical performance of these systems can be anticipated.

It is worth noting that for the Mo- based monolayers, the dispersion of valence valleys at the K point changes insignificantly when going towards heavier Z atoms, but the dispersion of conduction valleys at K drastically increases from N to Sb. As mentioned above, for all Z except N, the CBM+1 valley at K moves downward towards CBM for heavier Z atoms. This shift is accommodated by increasing the dispersion of CBM valley. As a result, the larger is the shift of CBM+1 valley, the larger is the dispersion of CBM at K. However, for W-based MLs, the change in the dispersion of CBM valley is only prominent in case of ML $WSi_2Sb_4$. This discrepancy can be understood by observing the splitting of CBM valley between K–Γ and K–M. This splitting increases with heavier Z elements and protects the dispersion CBM valley by deforming dispersion of CBM+1 valley. For ML $WSi_2Sb_4$, due to large splitting of both CBM and CBM+1 valleys, the dispersion of both valleys changes significantly.

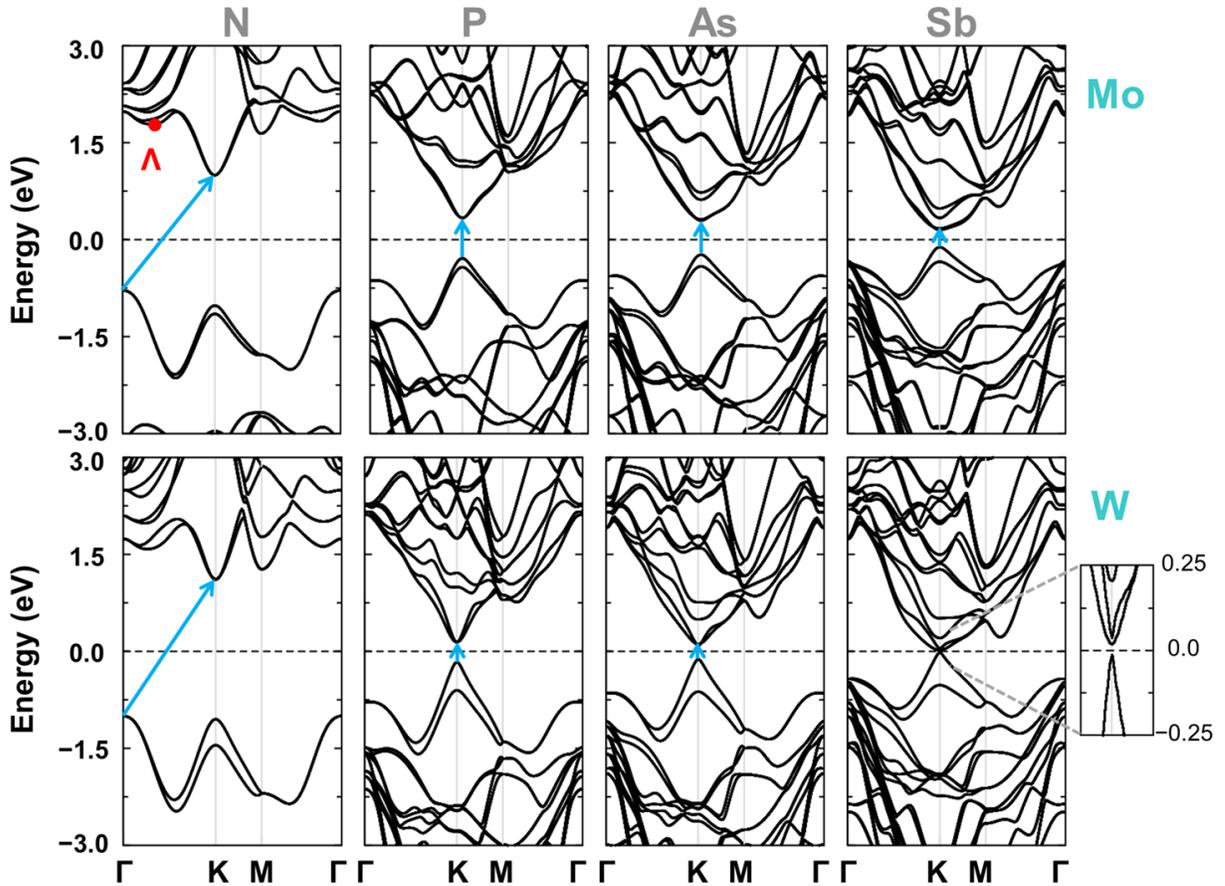

**Figure 2.** Band structures of $MSi_2Z_4$ MLs calculated at the PBE level of theory including spin-orbit coupling for Mo (top panel) and W (bottom panel) based systems. The Fermi level, set in the mid-gap to zero, is marked by a dashed line. The fundamental gap is indicated by a blue arrow. In the upper left most panel, Λ valley is marked.

Next, the elemental composition of frontier bands was analyzed for an exemplary system (ML $WSi_2P_4$) which revealed that VBM and CBM are mainly composed of different orbitals of W (M-) atoms, see **Figure S4**. The isosurface of second highest valley (VBM -1) at Γ point is mainly composed by $dz^2$ orbitals of W atoms with a considerable portion of isosurface on the A–Z bond positioned along the layer thickness. Changing Z atom increases the layer thickness and results in different overlap of these two isosurfaces. This different extent of overlap moves Γ valley on the energy scale[44,45], which controls the size or nature of the band gap. Surprisingly, the outer layers have no contribution in forming the frontier states, which may be the cause that its band edges are well protected by the local environment.[31]

The trend of band gap size, the frontier band dispersion, and the spin-splitting remain unchanged within GW approximation, and thus, $WSi_2N_4$ ($WSi_2Sb_4$) has the largest (smallest) QP gap, see Table 1. The only qualitative change occurs in the electronic structure of $WSi_2N_4$, where VBM shifts from Γ to K point, hence, turning it into direct gap semiconductor with the band gap of 3.05 eV. A similar indirect-direct transition for ML $WSi_2N_4$ was also reported when band structure was corrected with hybrid (HSE06) functionals.[36] However, ML $MoSi_2N_4$ remains indirect within GW approximation, but the energy difference between valence band maxima at K and Γ points reduces to 139 meV. Since these values are sensitive to the convergence parameters of the GW calculations, we checked them carefully, as reported in the SI.

### iii. Optical absorption spectrum and exciton wave functions

To assess the optical performance of $MSi_2Z_4$ MLs, the optical absorption spectra were computed by solving the BSE, see results in **Figure 3** and **Table 1**. The N-containing MLs are characterized by the first excitation in the visible range, see **Figure 3**a and e, while other ones have their onset in the IR region, see **Figure 3**. The lowest energy excitons in all materials stem from the transition between the VBM and the CBM, see detailed analysis in the SI, **FigureS7** and **FigureS8**.

Above the absorption onset, all materials exhibit large absorption with pronounced maxima at the upper boundary of the visible region (see Figure 3). Interestingly, these features overlap well with the lowest-energy peaks in the spectra of the N-containing MLs, suggesting the possibility of stacking them with other, heavier members of this material family (e.g., containing P or As) to form tandem solar cells. To check the accuracy of our calculations, we compared the calculated energies of the 1st and 2nd excitations with the available experimental data.[28] Hong et al.,[28] reported the 1st and 2nd peaks of $MoSi_2N_4$ at 2.210 and 2.350 eV, respectively, corresponding to the two direct excitations originating from the VBM splitting of 140 meV. In our calculations, the 1st and 2nd excitation peaks of $MoSi_2N_4$ are at 2.470 and 2.618 eV which are slightly blue-shifted and exhibit a similar splitting to the experimental one. For reference, the VBM splitting of $MoSi_2N_4$ at the PBE level is 130 meV.

The energies of the first excitation and the binding energies decrease by increasing the mass of Z, while the oscillator strength follows the opposite trend. In general, the W-based MLs exhibit excitations with larger intensities than their Mo-based counterparts. The lowest-energy excitons of WSi$_2$N$_4$ and MoSi$_2$N$_4$ (WSi$_2$Sb$_4$ and MoSi$_2$Sb$_4$) have the largest (smallest) binding energies, equal to 450 and 390 meV (160 and 120 meV), respectively. Regardless of the larger thickness of MA$_2$Z$_4$, their binding energies and band gaps follow the linear scaling law, as previously reported for conventional 2D materials, see **Figure 4**a[46]. It should be noted that the values of the binding energies are highly sensitive to the convergence parameters employed in the GW and BSE calculations. For a thorough discussion in this regard, see **Figure S2-3** and **Figure S5-6** for the convergence of GW band gap and binding energy.

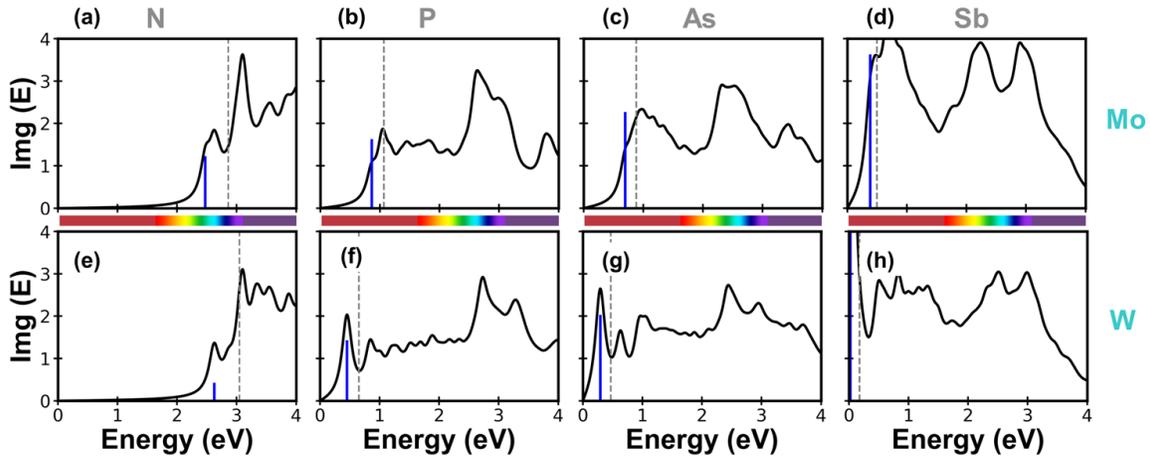

**Figure 3.** Optical absorption spectra of the considered MSi$_2$Z$_4$ MLs. The vertical dashed lines indicate the GW band gaps and the solid blue bars, the first excitation. For reference, the visible light with IR and UV range is marked in the middle.

In addition to the usual analysis of the exciton binding energies *versus* the QP gap, we inspected the trends for E$_b$ with respect to the static dielectric screening calculated from the random-phase approximation entering the solution of the BSE.[47,48] From the results plotted in **Figure 4**b, we notice that, contrary to the intuition based on the number of electrons in the metal atoms, W-containing MLs feature a lower screening than their Mo-based counterparts. We assign this behavior to the structural characteristics of the MSi$_2$Z$_4$. As shown in **Table 1**, the thickness of the MLs ranges from 7.0 to 10.9 Å, for going towards heavier Z elements, but it does not depend on the M atoms, which are buried in the inner layer. Due to the same thickness of MLs with same non-metallic, MoSi$_2$Z$_4$ are, thus, more compact than WSi$_2$Z$_4$ and, as such, feature larger polarizability.

Examining now the results shown in **Figure 4**b, we find that the biding energies decrease with increasing size of the Z element, upon which its layer thickness becomes larger and the QP gap smaller. The approximately linear behavior identified with respect to the gap in **Figure 4**a is not reproduced for the static screening. However, the trend shown in **Figure 4**b suggests closer similarities among the same non-metallic compositions (MoSi$_2$Z$_4$ and WSi$_2$Z$_4$) at varying Z than among materials with the same metal element. Again, this finding is promising in view of constructing heterostructures based on these systems: by choosing equal non-metallic compositions, which also exhibit negligible lattice mismatch as discussed above, it can lead to a bilayer in which approximately the same amount of energy is required to dissociate the excitons in either side of the interface. This prediction holds particularly for P- and As-containing materials.

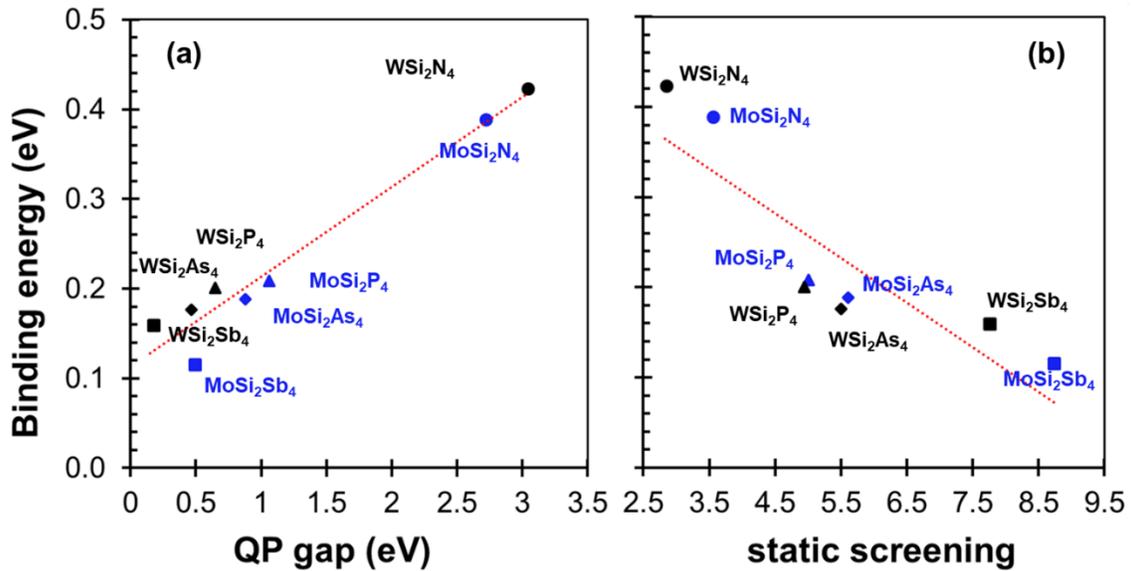

**Figure 4.** Binding energies of the lowest-energy exciton plotted with respect to a) the QP gap and b) the static screening. The linear fitting is indicated by the red dashed line.

## 3.1 Summary and Conclusions

In summary, we have investigated with first-principles many-body calculations the quasi-particle electronic structure and the optical properties of $MSi_2Z_4$ monolayers with M = Mo, W and Z = N, P, As, Sb. All systems have a fundamental direct band gap with the valence band maximum (VBM) and conduction band minimum (CBM) at the K point ranging from about 3 eV in $WSi_2N_4$ down to about 0.15 eV in $WSi_2Sb_4$. $MoSi_2N_4$ features an indirect QP gap of approximately 2.7 eV with the VBM at Γ: it differs by about 200 meV from the optical gap. Upon incrementing the mass of Z, the spin-orbit splitting increases leading to a decrease of the band gap. $WSi_2N_4$ has the largest band gap and $WSi_2Sb_4$ offers the smallest band gap. Unlike conventional TMDCs, $MSi_2Z_4$ do not exhibit the Λ valley in the conduction region, which is also known as channel for nonradiative recombination, quenching photoluminescence efficiency. The considered materials have an intriguing composition-dependent absorption. The spectra of $MoSi_2N_4$ and $WSi_2N_4$ have their absorption onset in the visible region, while all the other materials absorb IR radiation. All monolayers exhibit intense excitations at the onset indicating they are good light absorbers. The first exciton stems from the transition between the VBM and CBM meaning that electron and hole dissociated from the exciton will be localized. Exciton binding energies range from 0.42 eV in $WSi_2N_4$ to 0.12 eV in $MoSi_2Sb_4$, much lower than these of corresponding TMDC MLs. In general, N- (Sb-) based monolayers have the largest (smallest) binding energy, which decreases linearly when changing the Z element from N to Sb, following a linear scaling law between the band gaps and binding energies in 2D materials[46]. By contrasting binding energies with the calculated value of the static screening, one does not notice an equally clear trend. Nonetheless, important findings include the larger dielectric screening of the Mo-based MLs compared to the W-based ones, due to the more compact structure of the latter, as well as the larger similarity exhibited by materials differing only by the metal atom.

The results of this study indicate $MSi_2Z_4$ as a family of favorable light absorbers at energies that vary with their composition. The systems with Z = P, As, Sb, absorbing IR radiation, can be favorably combined with other low-dimensional semiconductors, such as conventional TDMCs or even $MSi_2N_4$, that absorb in the visible range to form tandem solar cells. The combination of several MSi$_2$Z$_4$ layers, also with different compositions, is particularly attractive due to the lattice parameter independent of the metal atom for a given combination or A and Z atoms. Dedicated studies exploring these heterostructures are needed to assess their behavior, but our results provide a suitable basis to conduct this analysis.

## 4.1   Computational details

All calculations presented in this work were performed within the framework of DFT [49] and many-body perturbation theory[47] as implemented in the Vienna ab-initio simulation (VASP) code [50]. The interactions between electrons and nuclei in the Kohn-Sham (KS) equations [51] were treated with the projector augmented wave (PAW) method [52]. In all calculations, the generalized gradient approximation for the exchange correlational potential as proposed by Perdew, Burke, and Ernzerhof (PBE) [53] was employed, along with Grimme's DFT–D3 dispersion correction [54] to account for the contributions of van der Waals (vdW) interactions. Spin-orbit coupling (SOC) was included in all calculations. To minimize spurious interactions between periodic images, a vacuum of ~15 Å was inserted along the non-periodic lattice vector. The unit-cell parameters and the atomic positions were optimized using a Γ-centered 12×12×1 **k**-point mesh with a plane wave energy cutoff of 500 eV. The structures were optimized until the residual interatomic forces were less than 10 meV Å$^{-1}$, with an electronic self-consistency convergence threshold of 10$^{-8}$ eV. Crystal structures were visualized using VESTA [55].

The optical properties were calculated from the GW approximation and the solution of the Bethe-Salpeter equation (BSE). The quasiparticle eigenvalues were obtained starting from the PBE energies and wave functions by solving the quasiparticle (QP) equation[29]:

$$[T + V_{ext} + V_H + \Sigma(E_m^{QP})]\psi_m^{QP} = E_m^{QP}\psi_m^{QP}, \quad (1)$$

where the self-energy operator $\Sigma$ is calculated in the single-shot flavor (G$_0$W$_0$) of the GW approximation[56]. The other terms in Eq. (1) are the kinetic energy (*T*), the external potential accounting for the electron-nuclear attraction ($V_{ext}$), and the Hartree potential ($V_H$); $E_m^{QP}$ and $\psi_m^{QP}$ are the single-particle energies and the wave functions corrected with the self-energy contribution. Optical absorption spectra were obtained by solving the BSE, the equation of motion of the two-particle correlation function[57], on top of the QP electronic structure. In practice, the problem is cast into a Schrödinger-like equation with the form:

$$\left(E_{ck}^{QP} - E_{vk}^{QP}\right)A_{vck}^S + \sum_{c'v'} K_{(e-h)vck,v'c'k}(\Omega^S)A_{c'v'k}^S = \Omega^S A_{vck}^S, \qquad (2)$$

where $A_{vck}^S$ are exciton amplitudes, $K_{(e-h)}$ is the kernel describing the electron-hole (e-h) interactions, and $\Omega^S$ are the excitation energies. The coefficients, $A_{vck}^S$, provide information about the single-particle transitions contributing to the exciton. They can be visualized in the reciprocal space using the so-called exciton weights defined as:[58]

$$w_{vk}^S = \sum_c |A_{vck}^S|, \qquad w_{ck}^S = \sum_v |A_{vck}^S|$$

where $S$ is the index of the exciton.

In the GW calculations, we chose a total of 240 bands, a cutoff energy of 100 eV, 100 frequency points, and a Gaussian smearing of 50 meV. The **k**-point mesh to sample the Brillouin zone was doubled with respect to the choice adopted for the DFT calculations (24×24×1), except for WSi$_2$Sb$_4$ where an 18×18×1 mesh was used. A plane wave energy cutoff of 400 eV was adopted in these calculations. A set of 4 valence bands and 8 conduction bands was used to construct and solve the BSE.

## 5.1    Data Availability

The data to reproduce the plots and findings within this paper are available from the corresponding author(s) upon reasonable request.

## 6.1    Conflict Of Interests

The authors declare no competing financial or non-financial interests

## 7.1 Acknowledgements

M.S.R and C.C. thank the funding by the Lower Saxony Ministry of Science and Culture (programs Professorinnen für Niedersachsen and "Digitalization in the natural sciences", project SMART), by the QuanterERA II European Union's Horizon 2020 research and innovation programme under the EQUAISE project, Grant Agreement No. 101017733, and by the Federal Ministry for Education and Research (Professorinnenprogramm III). The computational resources were provided by the high-performance computing center of ZIH Dresden and by the by the North German Computer Alliance (project nip00063). M.S.R., T.W., and A.K. thank the Deutsche Forschungsgemeinschaft (project GRK 2247/1 (QM3) and project CRC1415, number 417590517) for financial support and the high-performance computing center of ZIH Dresden for computational resources. A.K. also acknowledges association with priority program (project SPP2244 (2DMP)). T.W. also acknowledges the financial support of National Science Centre, Poland within Project No. 2021/41/N/ST3/04516

Supporting Information

# Composition Dependent Absorption of Radiation in semiconducting MSi$_2$Z$_4$ Monolayers

Muhammad Sufyan Ramzan,[1*] Tomasz Woźniak,[2,3] Agnieszka Kuc,[3]

Caterina Cocchi[1,4]

[1]Institut für Physik, Carl von Ossietzky Universität, 26129 Oldenburg, Germany.

[2]Department of Semiconductor Materials Engineering, Wrocław University of Science and Technology, Wrocław, 50-370 Poland

[3]Helmholtz-Zentrum Dresden-Rossendorf, Center for Advanced System Understanding, CASUS, Untermarkt 20, 02826 Görlitz Germany.

[4]Center for Nanoscale Dynamics (CeNaD), Carl von Ossietzky Universität, 26129, Oldenburg, Germany.

Correspondence to: muhammad.sufyan.ramzan@uni-oldenburg.de, caterina.cocchi@uni-oldenburg.de

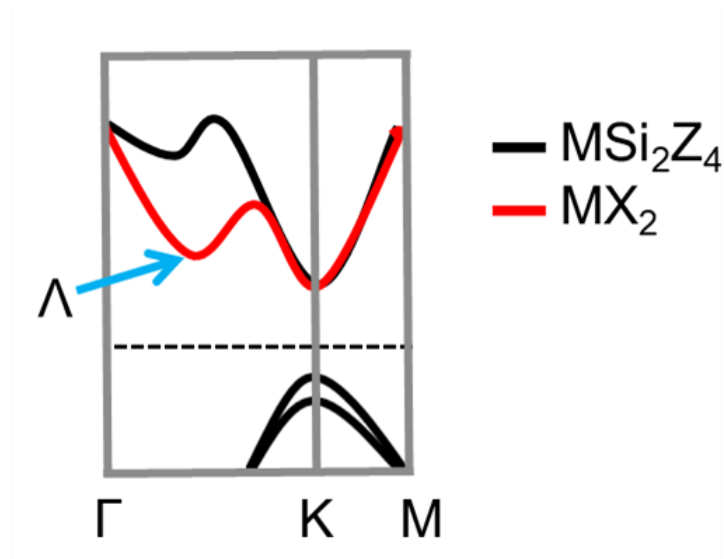

**Figure S5**. A schematic to show the presence (absence) of Λ valley in $MX_2$ ($MSi_2Z_4$) near the conduction band minimum.

### i. Convergence of GW Calculations

Due to lower accuracy of PBE functional, the band gaps were corrected with the single shot $G_0W_0$ calculations. It is worth mention that getting reliable values of GW band gaps for 2D materials is a difficult task due to its dependence of several convergence parameters, especially k-point mesh.[52–56] To get the reliable values of band gap, using $WSi_2N_4$ as a representative case, we check convergence of GW band gap with respect to k-point mesh and Coulomb truncation, see **Figure S1 and S2**. Our convergence tests show that GW band gap varies between 150 – 250 meV and band gaps calculated in this study matches well with earlier theoretical studies.[29,36]

We found that the optical band gap, estimated as the energy of the first bright excitation, converges much faster than the GW band gap,[57] especially with respect to vacuum, see **Figure S2 and S4** and **Figure S5** for the convergence of binding energy.

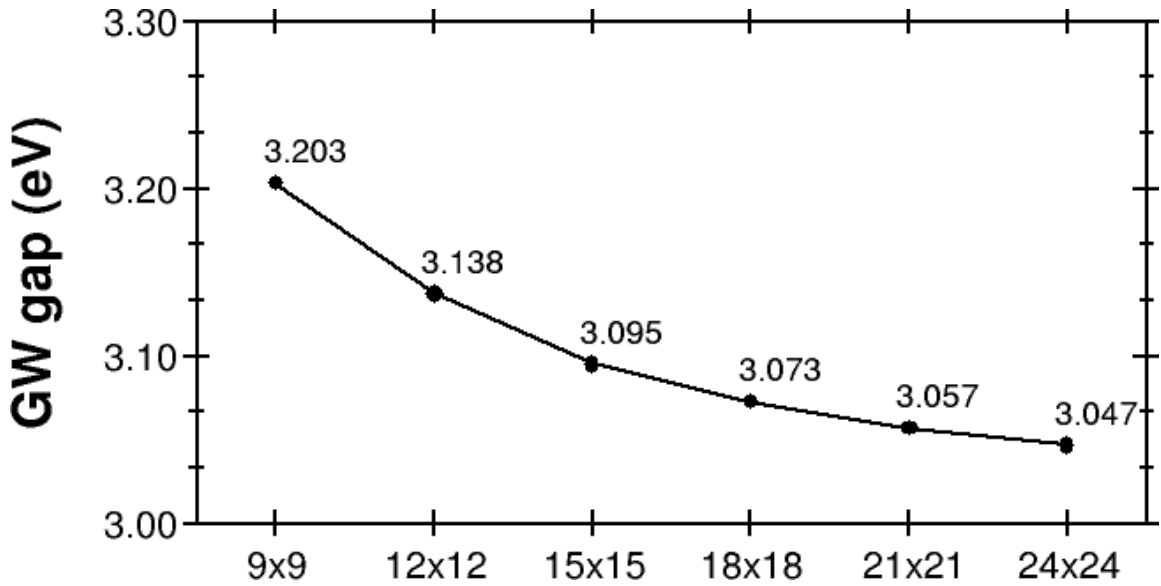

**Figure S6**. Convergence of GW band gap versus the k-point mesh using 400eV energy cutoff (ENCUT) and 240 total number of bands (NBANDS) for $WSi_2N_4$ monolayer. The overall change in the size of GW band gap is in the range of 200 meV and it changes only by 10 meV by increasing k-point mesh from 21x21 to 24x24.

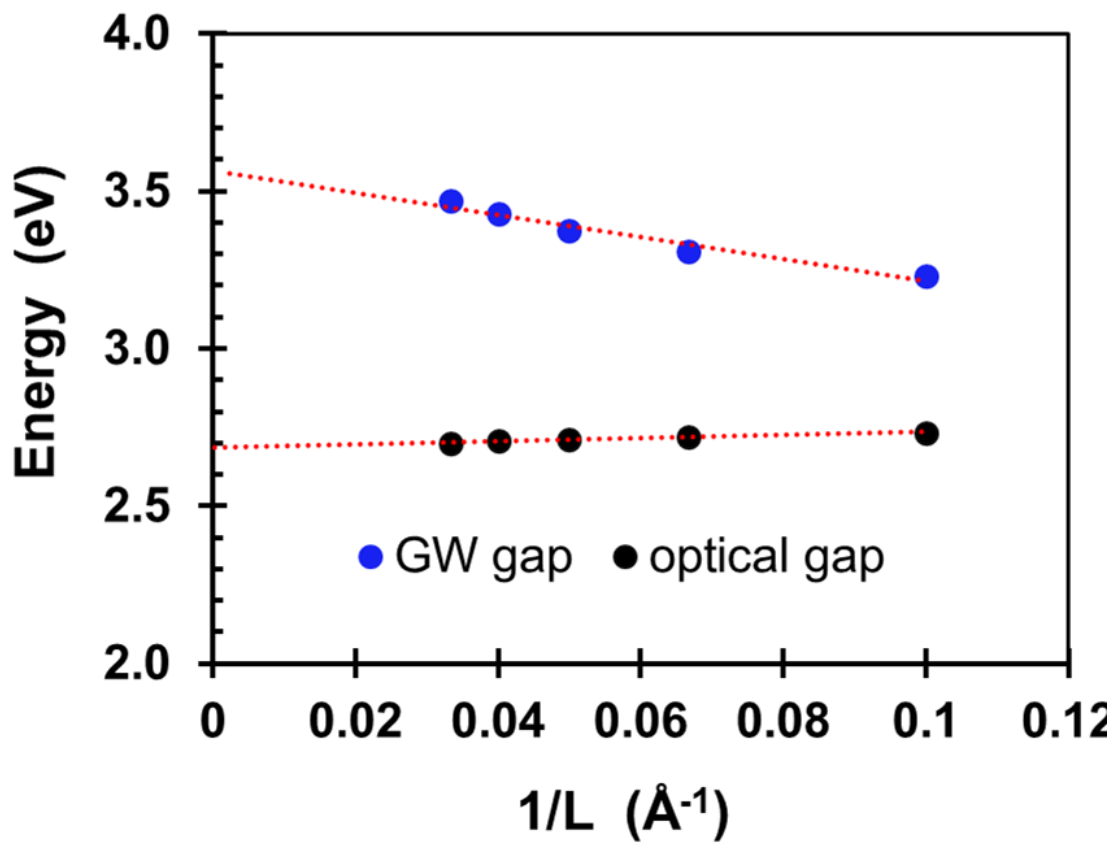

**Figure S7** Convergence of optical and GW band gaps w.r.t inverse of vacuum using 12x12x1 k-point mesh along with 400eV energy cutoff (ENCUT) and 240 total number of bands (NBANDS) for $WSi_2N_4$ monolayer. The red dotted line shows the linear fitting used to estimate values of band gap with infinite vacuum.

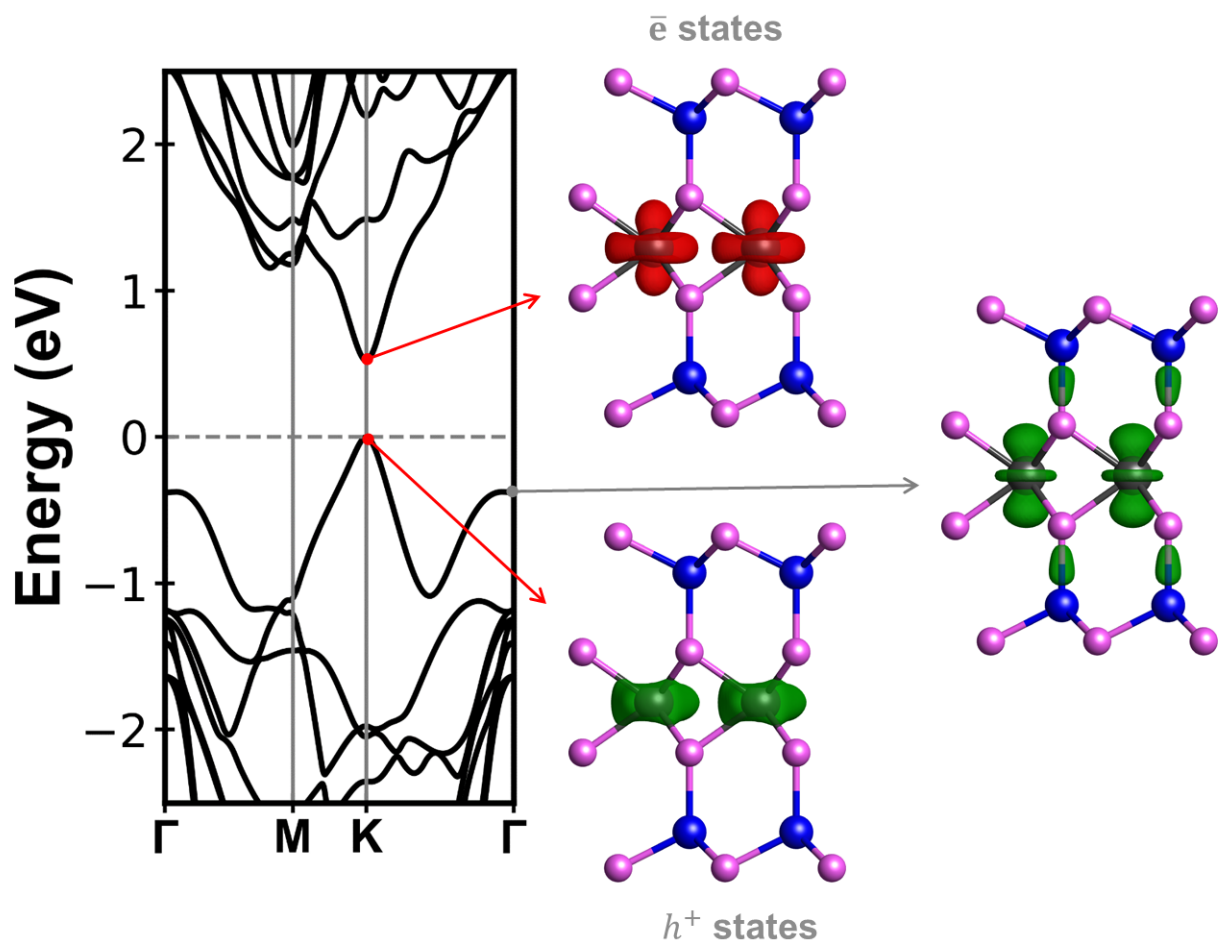

**Figure S8**. Electronic band structure (left panel) of ML WSi$_2$P$_4$ monolayer at the PBE level without SOC and real space wavefunctions at the high symmetry points ***K*** and **Γ** for valence band maximum (hole states: green isosurface) and conduction band minimum (electron states: red isosurface). The value of isosurface is set to 0.015e/A$^3$.

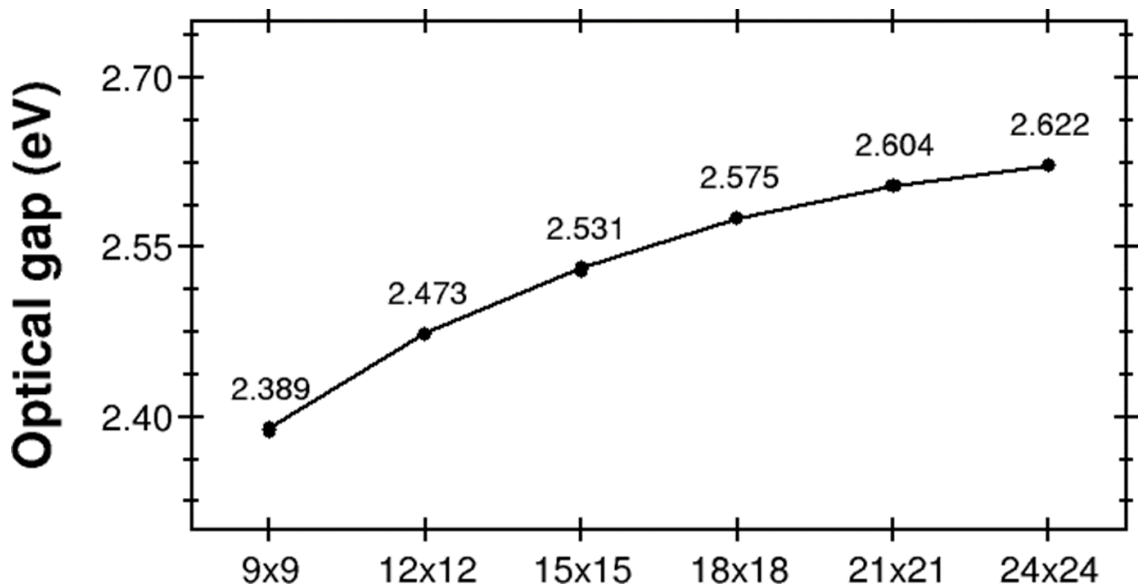

**Figure S9**. Convergence of optical gap versus the k-point mesh using 400eV energy cutoff (ENCUT) and 240 total number of bands (NBANDS) for $WSi_2N_4$ monolayer. The overall change in the size of optical band gap is in the range of 250 meV and it changes only by 18 meV by increasing k-point mesh from 21x21 to 24x24.

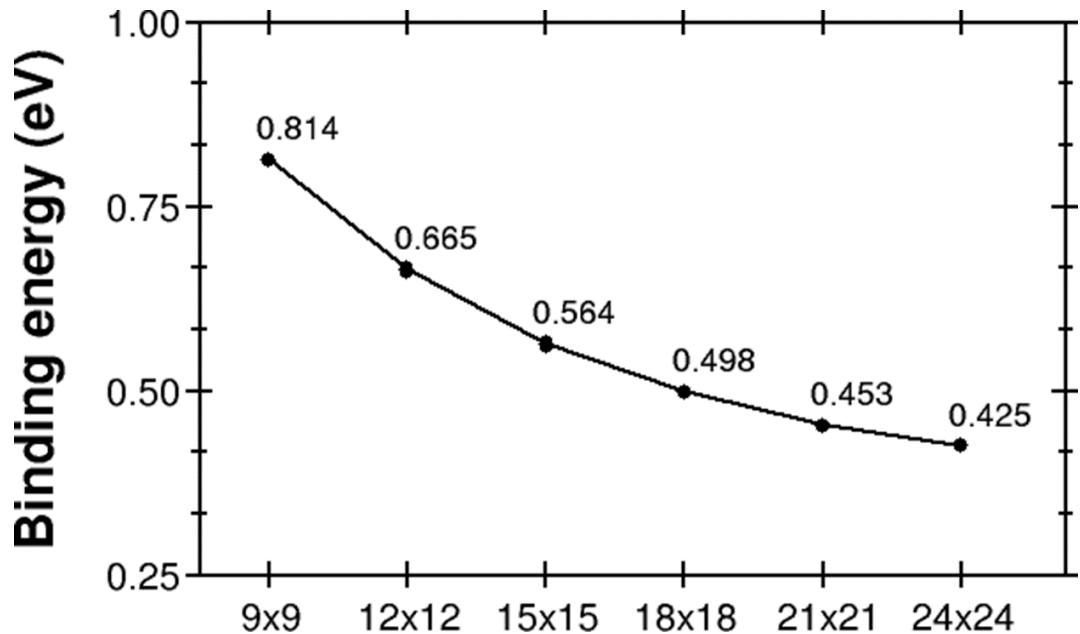

**Figure S10**. Convergence of binding energy versus the k-point mesh using 400eV energy cutoff (ENCUT) and 240 total number of bands (NBANDS) for $WSi_2N_4$ monolayer. The overall change in the value of binding energy is in the range of 400 meV and it changes only by 28 meV from increasing k-point mesh from 21x21 to 24x24.

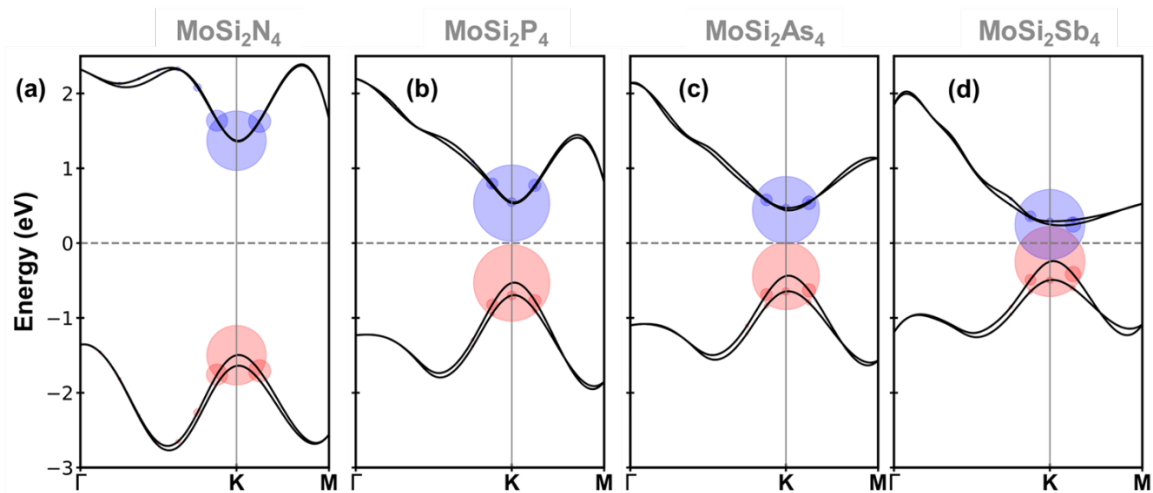

Figure S11 Distribution of the exciton coefficients on the GW band structure of the Mo-based monolayers: the radius of the colored dots indicates the relative weight of the corresponding electronic state. The horizontal dashed line indicates the Fermi level set in the mid-gap.

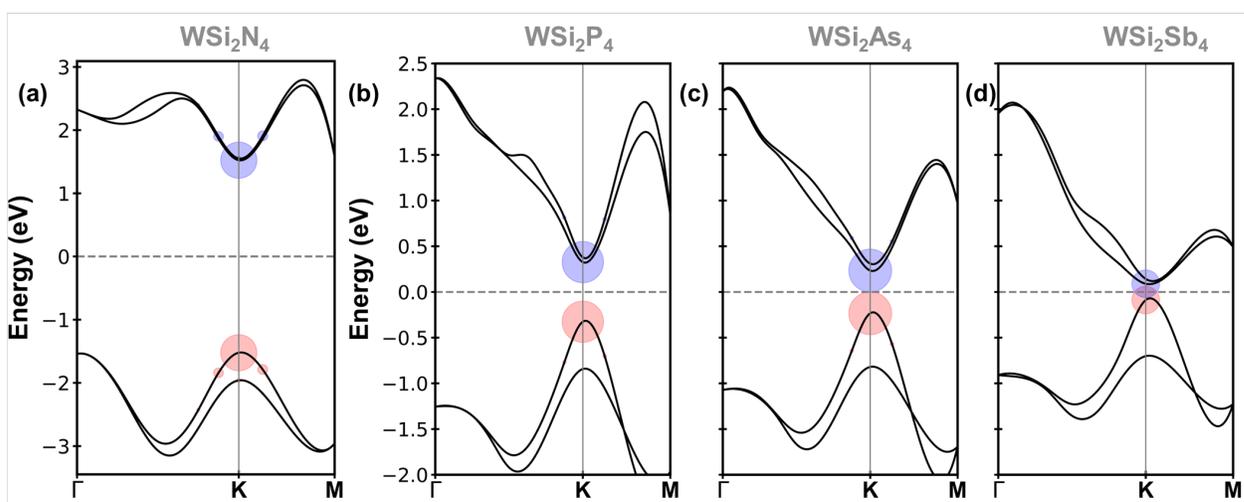

**Figure S12.** Distribution of the exciton coefficients on the GW band structure of the W-based monolayers: the radius of the colored dots indicates the relative weight of the corresponding electronic state. The horizontal dashed line indicates the Fermi level set in the mid-gap.

26